\documentclass[pra,showpacs]{revtex4}
\newcommand{\be}{\begin{equation}}
\newcommand{\ee}{\end{equation}}
\newcommand{\bea}{\begin{eqnarray}}
\newcommand{\eea}{\end{eqnarray}}
\usepackage{epsfig}
\usepackage{graphicx}

\begin{document}

 \title{Localization of an impurity particle on a boson Mott insulator background}

\author{Sevilay Sevin\c{c}li}
\author{R. O. Umucal{\i}lar}
\author{M.~\"O.~Oktel}
\email{oktel@fen.bilkent.edu.tr}
\affiliation{ Department of Physics, Bilkent University, 06800 Ankara, Turkey }%

\date{\today}

\begin{abstract}
We investigate the behavior of a single particle hopping on a
three dimensional cubic optical lattice in the presence of
a Mott insulator of bosons in the same lattice. We calculate the
critical interaction strength between the impurity and background
bosons, beyond which there is bound state (polaron) formation. We
give exact results in the limit of a perfect Mott insulator, where
polaron formation is equivalent to impurity localization. We
calculate the effects of lattice anisotropy, higher impurity bands,
and fluctuations of the Mott insulator on the localization
threshold. We argue that our results can be checked experimentally by RF spectroscopy of impurity particles.

\end{abstract}

 \maketitle

\section{Introduction}

Recent experiments using cold atoms in optical lattices have
displayed remarkable versatility, creating highly controllable,
isolated, low temperature environments where ideas about many
particle quantum mechanics can be tested \cite{greiner,bloch1,ketterle,esslinger,phillips,weiss}. With improvement over
the control of system parameters, as well as advancement of novel
measurement techniques such as noise correlations \cite{altman}, it seems
plausible that a great variety of models will be realized in
optical lattices. The precision of optical lattice experiments
presents another challenge for theory. Measurements of effects
that are beyond the simplest descriptions such as mean field theory, are becoming possible \cite{bloch2}.

One of the new classes of quantum models realized by cold gas
experiments is the mixture of different species of atoms \cite{jin}.
Boson-boson mixtures as well as boson-fermion mixtures have been
created. It is now also possible to selectively turn on optical
lattice potentials for any of the  species forming the mixture \cite{inguscio}.

The possibility of experimental realization of atomic gas mixtures stimulated a lot of theoretical interest \cite{molmer,albus,lewenstein,sacha,pethick,kuklov,gavish}. From the theory point of view, mixtures are
appealing as they can display some qualitatively new physics
stemming from simple ideas in many body theory. There is
possibility of pairing due to mediated interactions \cite{pethick}, formation of
composite particles similar to molecules \cite{lewenstein}, large counterflows of
different species or even countersuperfluidity \cite{kuklov}. There are also
ideas to simulate random potentials using one species as the
disorder potential for the others \cite{gavish}. Two recent experiments about
boson-fermion mixtures have shown that the optical lattice
experiments are advanced enough so that all these theoretical
ideas may now be tested in the laboratory \cite{gunter,ospelkaus}.

The parameter space for  mixtures is very large, with many
possible phases and a complicated phase diagram \cite{lewenstein,sacha}. The phase diagram
of mixtures changes remarkably depending on the strength and
type of optical lattices,  number of components, nature of
interactions, and the identity of the particles. While more
general investigations of this parameter space, which point out
novel phases, are very useful, in this paper, we concentrate on
a simple limit and provide some exact results. Experimentally, it
would be easy to check such results and gain a better
understanding of the various effects that may be present in the
system.

We investigate the limit where one bosonic species interacts
with a single particle of another species. While the identity of
the external particle does not matter in what follows, we refer to it as the fermion for convenience. All the results of the
paper are valid for a bosonic impurity as well. We assume that
both species share the same lattice potential and that the lattice
potential is deep enough so that only one band of the lattice is
populated. The Hubbard-type Hamiltonian \cite{jaksch} for this system can be written as
\be
 H=-t_b\sum_{< i,j>}b_i^\dagger b_j+\frac{U_{bb}}{2}\sum_in_i(n_i-1)-\mu_b\sum_in_i-t_f\sum_{< i,j>}f_i^\dagger f_j+U_{bf}\sum_in_if_i^\dagger
 f_i,
\ee
 where $b^\dagger, b$ and $f^\dagger, f$ are the bosonic and
impurity creation and annihilation operators respectively;
$n_i=b_i^\dagger b_i$ is the on-site number operator for the
bosons and $<i,j>$ represents a sum over nearest neighbors. The
strength of the tunnelling terms are characterized by hopping
matrix elements $t_b$ for bosons and $t_f$ for the fermion. $U_{bb}$ and $U_{bf}$ are the on-site interaction strengths
between bosons and between a boson and a fermion respectively.

Depending on the ratio between the the tunnelling strength and the
interaction for bosons, the system may either form a superfluid
(SF) or a Mott insulator (MI) state \cite{fisher}. While the
localization problem can be studied in both regimes, we can
provide exact results only for the Mott insulator case. In this
paper, we concentrate on the problem in the Mott regime as the
problem is qualitatively different from that of a superfluid
interacting with an impurity. Thus, we consider $t_b/U_{bb} \ll
1$, give exact results only in the limit $t_b/U_{bb} = 0$ and
calculate corrections
 due to finite boson hopping. For a Mott insulator with $n_0$ bosons per site, the chemical potential is constrained to
 \be
 U_{bb} (n_0-1) \le \mu_b \le U_{bb} n_0.
 \ee

 With these considerations for bosons, the fermion can show two qualitatively different behaviors.
 It can either behave like a free fermion with its wavefunction stretching throughout the system, or it may create
 a defect in the Mott insulator and form a bound state with this defect. We calculate the critical interaction strength that separate these two regimes. In Section \ref{sec2}, we study the case of a perfect Mott insulator and obtain the phase diagram for localization of the impurity; in Section \ref{sec3}, we generalize this exact result to an anisotropic lattice
 where hopping strength in one direction is different from the other two. We consider the effect
 of higher impurity bands in Section \ref{sec4}. In Section \ref{sec5}, we calculate the change in the fermion hopping
 strength due to fluctuations in the Mott insulator background. We summarize our
 results and discuss experimental methods to measure this bound state formation in Section \ref{sec6}.

\section{Localization in a perfect Mott insulator}\label{sec2}

In this section, we calculate the critical interaction strength for bound state (polaron)
formation in the limit that the boson Mott insulator is perfect,
i.e. the hopping strength for bosons is zero ($t_b/U_{bb}=0$). We
relax this assumption in Section \ref{sec5}, and analyze the effects of
fluctuations in the Mott background. When boson hopping is neglected the Mott insulator background
becomes almost inert for the fermion, presenting a spatially
independent mean field energy shift. In this case the fermion will
move with the dispersion relation
\be
E_k = t_f \bigg[ 6 -2\sum_{i=x,y,z}\cos(k_ia) \bigg],
\ee
where $-\pi/a < k_i \le\pi/a$ is the crystal momentum of the fermion in the $i$ direction
and $a$ is the lattice constant. We cannot expect this delocalized
behavior of the fermion to continue if the interactions between
the background bosons and the fermion become very strong. Let us
assume that the fermion and the bosons attract each other, i.e.
$U_{bf} < 0$. In this case, it would be energetically favorable to
put more bosons at a lattice site and bind the fermion to these
bosons. This will only happen at a critical interaction strength,
beyond which the energy gained by boson-fermion attraction is
larger than the sum of the kinetic energy cost of localizing the
fermion and the interaction energy cost of introducing more
bosons.

This critical interaction strength can then be found by
investigating the single particle Hamiltonian
\begin{equation}\label{hamilton}
 H=-t_f\sum_{< i,j>}f_i^\dagger f_j-Vf_0^\dagger f_0,
\end{equation}
where we take the site at which the defect is formed as the origin
and assume that the defect represents a localized attractive
potential $-V$ to the fermion. We can then ask for which value of
$V$ there will be a bound state in the spectrum. This is the
discrete version of the problem of existence of a bound state
for a localized potential well \cite{landau}. Just like the continuum version, in
one and two dimensional lattices there is a bound state for an
infinitesimally small attractive potential. In three dimensions, however,
there is a certain critical value below which there is no bound
state. In the case of a finite attractive potential, the dispersion
relation takes the following implicit form
\begin{equation}\label{auxiliary}
\phi_k\big[\tilde{E} +2\sum_i\cos(k_ia)\big] = -\tilde{V} \psi_0,
\end{equation}
where $\tilde{E}=E/t_f$ and $\tilde{V}=V/t_f$ are the scaled
quantities and $\phi_k=\sum_i \psi_i\mathrm{e}
^{-i\mathbf{k}\cdot\mathbf{r_i}}$ is the Fourier transform of the particle's wave function. We obtain the
relation between the binding energy and the attractive potential
by taking the Fourier transform of Eq. (\ref{auxiliary})
\begin{equation}\label{v1int}
 \frac{1}{\tilde{V}}=\int_{-\pi}^\pi\int_{-\pi}^\pi\int_{-\pi}^\pi\frac{d\theta_xd\theta_yd\theta_z}{(2\pi)^3}\frac{1}{(6-2\sum_i\cos
 \theta_i)+\epsilon},
\end{equation}
where $\theta_i=k_ia$ and we take
\begin{equation}
 \tilde{E}=-6-\epsilon,
\end{equation}
with $\epsilon >0$ being the binding energy. When $\epsilon=0$, i.e.
at the localization threshold, the above integral can be evaluated exactly \cite{delves} and the critical value at which the
localization takes place is
\begin{equation}
 \tilde{V_c}=\frac{2}{(18+12\sqrt{2}-10\sqrt{3}-7\sqrt{6})\left[\frac{2}{\pi}K(k_0)\right]^2}=3.95678,
\end{equation}
where $k_0^2=[(2-\sqrt{3})(\sqrt{3}-\sqrt{2})]^2$ and $K$ is
the complete elliptic integral of the first kind. For nonzero $\epsilon$, the integral was
evaluated by Joyce \cite{joyce}. Using this result, the potential is obtained as
\begin{equation}
 \tilde{V}=\frac{2}{\frac{(1-\eta)^{1/2}}{\omega}\left(1-\frac{1}{4}\eta\right)^{1/2}\left(\frac{2}{\pi}\right)^2
 K(k_+)K(k_-)},
\end{equation}
where $\eta=-16z(\sqrt{1-z}+\sqrt{1-9z})^{-2}$,
$z=1/\omega^2=1/(3+\epsilon/2)^2$, and $k_\pm^2=\frac{1}{2}\left[1\pm
\eta\sqrt{1-\frac{1}{4}\eta}-\left(1-\frac{1}{2}\eta\right)\sqrt{1-\eta}\right]$. In the limit of large binding energy, $\epsilon>>1$, we obtain a linear relation $\tilde{V}\propto\epsilon$, which is expected as the particle is strongly localized at a single lattice site.

\begin{figure}
 \centering
 \includegraphics[scale=0.8]{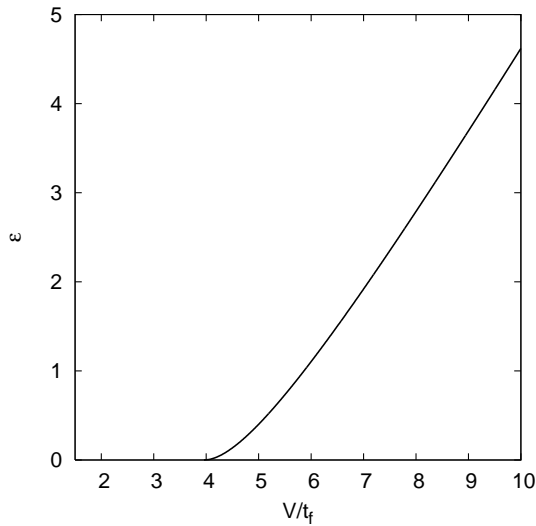}
  \caption{Binding energy $\epsilon$ of the impurity as a function of $V/t_f$. The critical interaction strength where the localization begins can also be obtained from the figure, i.e. $\epsilon=0$ for $V/t_f\lesssim 3.96$.}
 \label{fig:1}
\end{figure}

The exact evaluation of the integral above allows us to
calculate not only the critical boundary but also the
binding energy $\epsilon(V/t_f)$ of the bound state (Fig.
\ref{fig:1}).
We now translate our one particle results to the many particle
case. Let us first assume that
 the fermion-boson interaction is attractive, $U_{bf}<0$. In this case the simplest defect
 would be to introduce one more boson, thus the attractive potential seen by the fermion will be
 $V=|U_{bf}|$. However, to introduce one more boson to a Mott insulator with $n_0$ particles per lattice site would
 cost energy, $ U_{bb} n_0 - \mu$ . Thus, the phase boundary between the free fermion state and the bound state of the fermion and one bosonic defect (polaron) is given by
 \be
 \epsilon(\frac{-U_{bf}}{t_f})=\frac{U_{bb}n_0 - \mu }{t_f}.
\ee
This is by no means the only defect that can be created in the
Mott insulator. If the boson-fermion attraction is strong enough,
it becomes energetically favorable to attract more bosons and form
a bound state of two bosons and one fermion. The phase boundary
for such a defect can be decided by comparing the energy of this
state with the energy of the bound state of one boson and one
fermion. Thus, the equation for phase boundary is \be
\epsilon(\frac{-U_{bf}}{t_f})-\frac{U_{bb}n_0- \mu }{t_f} =
\epsilon(\frac{- 2 U_{bf}}{t_f})- \frac{U_{bb} (2 n_0 + 1) - 2
\mu}{t_f}. \ee One can similarly find the boundaries for bound
states with higher number of bosons.

\begin{figure}
 \centering
 \includegraphics[scale=0.8]{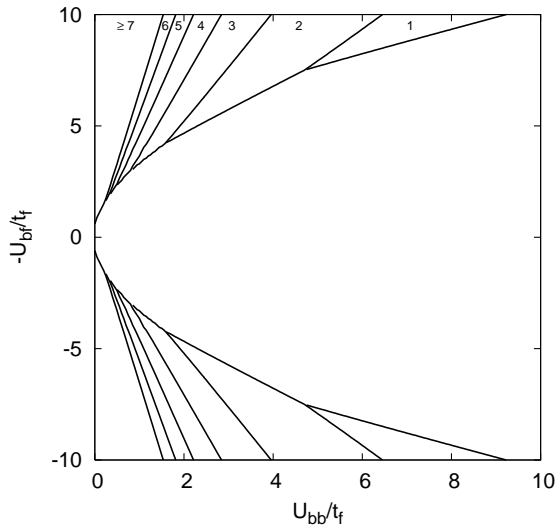}
  \caption{Phase diagram for $\mu = (n_0-1/2)U_{bb}$.
  Numbers in each region show how many extra particles ($U_{bf} < 0$) or holes ($U_{bf} > 0$)
  are attracted to the localization site. The region marked as $\geq 7$ contains all the phases
  with seven or more extra bosons (holes). Phase diagram for this value of $\mu$ is symmetric around
  $U_{bf}=0$. For small boson-boson repulsion $U_{bb}$, even for small $|U_{bf}|$ values, large number
  of bosons are attracted. While this phase diagram is independent of $n_0$, the number of holes that
  are attracted is limited by $n_0$.}
 \label{v_ubb_minener_01}
\end{figure}

\begin{figure}
 \centering
 \includegraphics[scale=0.8]{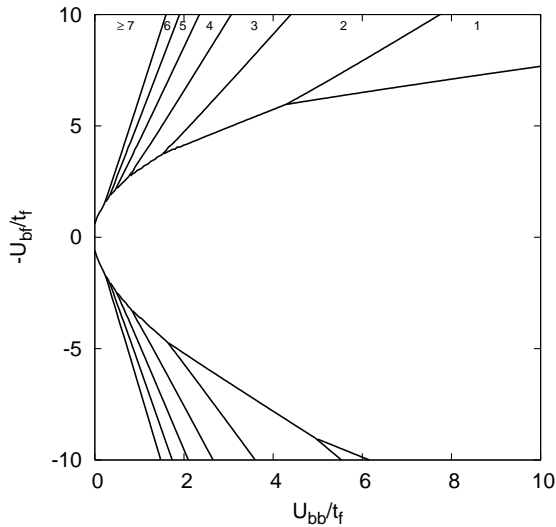}
  \caption{Phase diagram for $\mu = (n_0-1/4)U_{bb}$. Symmetry in Fig. \ref{v_ubb_minener_01} is broken and particle attraction is easier than the hole attraction, since the chemical potential is increased with respect to the symmetry point. To attract a hole one needs higher boson-fermion interaction $|U_{bf}|$ for the same $U_{bb}$.}
 \label{v_ubb_minener_01_mup}
\end{figure}

\begin{figure}
 \centering
 \includegraphics[scale=0.8]{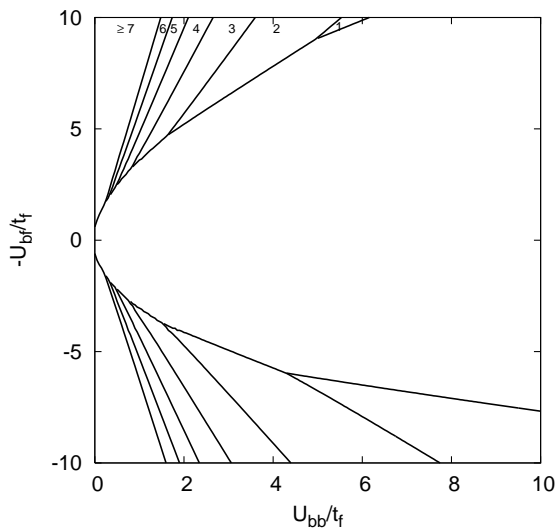}
  \caption{Phase diagram for $\mu = (n_0-3/4)U_{bb}$. As opposed to Fig. \ref{v_ubb_minener_01_mup}, to attract a particle one needs higher boson-fermion interaction.}
 \label{v_ubb_minener_01_mum}
\end{figure}

Another kind of defect would present itself for repulsive
interactions, $U_{bf}>0$. For sufficiently strong repulsive
interactions it would be preferable to create a hole in the Mott
insulator state and bind the fermion to this hole. The
corresponding phase boundary is given by
\be
\epsilon(\frac{U_{bf}}{t_f}) = \frac{- U_{bb}(n_0 - 1) +
\mu}{t_f}.
\ee
Similar to the attractive interactions, it is
possible to form bound states of the fermion with more holes. One
can continue to deplete the Mott state until all the $n_0$ bosons are removed from the defect site. After this point it
would be preferable to deplete bosons from the neighboring sites. We have, however, not included such states in our phase diagram.
In Figs. \ref{v_ubb_minener_01}, \ref{v_ubb_minener_01_mup}, and
\ref{v_ubb_minener_01_mum}, we present three phase diagrams for
different chemical potentials. Fig. \ref{v_ubb_minener_01}
indicates that when $\mu = (n_0-1/2)U_{bb}$, phase diagram is symmetric around
$U_{bf}=0$. This is expected as this value of $\mu$ corresponds to lobe centers of the Bose-Hubbard phase diagram where there is particle-hole symmetry. One can also notice from this
diagram that when $U_{bb}$ is close to zero, even for small
$|U_{bf}|$ values, the fermion can be bound to a large number of
bosons. If we take $\mu = (n_0-1/4)U_{bb}$ as in Fig.
\ref{v_ubb_minener_01_mup}, the symmetry around $U_{bf}=0$ is broken and for the repulsive interactions it is harder to attract holes. Fig. \ref{v_ubb_minener_01_mum} represents
the opposite case ($\mu = (n_0-3/4)U_{bb}$) where stronger interactions are required to attract particles.


We believe that the phase diagram can be checked experimentally. While it would be
possible to modify $U_{bf}$ by an interspecies Feshbach resonance,
an easier route would be to change $t_f$ which is controlled by
the strength of the optical lattice. The localized impurity states
can be distinguished from free fermion states as their mean
field shifts would be different; in principle, RF spectroscopy \cite{hadzibabic,grimm} would
directly detect the difference in the mean field shift. Although
the calculation was carried out for a single impurity, we expect these results to be quantitatively correct for a small
density of fermionic impurities over a bosonic Mott insulator
background. Essentially, if the inverse of Fermi momentum is much
larger than the lattice spacing, then the fermions would hardly
effect each other's behavior.

In this section, we obtained the phase diagram for the
interaction of one impurity particle with a perfect Mott
insulator. In the next three sections, we investigate how this
ideal situation is affected by lattice anisotropy, higher
impurity bands, and fluctuations of the Mott insulator.

\section{Effects of Lattice Anisotropy}\label{sec3}

In the optical lattice experiments, it is possible to change the
strength of the laser beams forming the lattice, hence realize a
model system where the lattice is not isotropic. For a
quantitative comparison with experiment, it
 is necessary to take this effect into account. We assume that the hopping strength
 for the fermion is different in one direction compared to the other two directions
 and calculate the effect of such anisotropy on the phase diagram of the previous section.

The localization threshold for the anisotropic case can also be
calculated analytically. Thus, in the following discussion we need
not assume that the anisotropy of the lattice is small. As a
simple limit, we obtain the two dimensional lattice localization
problem as the layers are decoupled. For that
limit, even the smallest attraction between the fermion and bosons
causes bound state formation. As the coupling between the two
dimensional layers is increased the critical interaction needed
for bound state formation monotonically increases.

\begin{figure}
 \centering
 \includegraphics[scale=0.8]{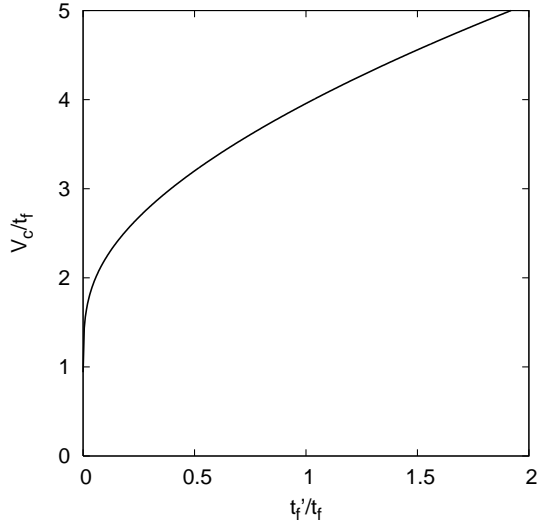}
  \caption{The critical value of interaction $V_c/t_f$ as a function of lattice anisotropy characterized by $t_f^\prime/t_f$ (Eq. (\ref{Vc2})). If the hopping parameter $t_f^\prime$ increases, anisotropy of the lattice increases and the localization becomes more difficult. $\tau=t_f^\prime/t_f=1$ gives $V_c/t_f$ for the isotropic case.}
 \label{fig:2}
\end{figure}

Because of the anisotropy, the single particle Hamiltonian in Eq.
(\ref{hamilton}) is modified as
\begin{equation}
 H=-t_f\sum_{< i,j>}f_i^\dagger f_j-t_f^\prime\sum_{< i,j>}f_{iz}^\dagger f_{jz}-Vf_0^\dagger
 f_0,
\end{equation}
where we take the hopping term in the $z$ direction to be $t_f^\prime\neq t_f$. We obtain the relation between $\tilde{V}$ and $\epsilon$ as
\begin{equation}
 \frac{1}{\tilde{V}}=\int_{-\pi}^\pi\int_{-\pi}^\pi\int_{-\pi}^\pi\frac{d\theta_xd\theta_yd\theta_z}{(2\pi)^3}\frac{1}{[4-2\sum_{i=x,y}\cos
 \theta_i+2\tau(1-\cos\theta_z)]+\epsilon},
\end{equation}
where $\tilde{E}=-4-2\tau-\epsilon$ and $\tau=t_f^\prime/t_f$. For $\epsilon=0$ this integral can be evaluated exactly \cite{delves}. The critical value for the
localization is found to be
\begin{equation}\label{Vc2}
  \tilde{V_c}=\frac{2}{\frac{\sqrt{2}}{\tau}(\sqrt{2}\sqrt{1+\tau}-\sqrt{2+\tau})\left(\frac{2}{\pi}\right)^2K[k_+(\tau)]K[k_-(\tau)]},
\end{equation}
where
\begin{equation}
 k_\pm(\tau)^2=\left[\frac{1}{\tau}(\sqrt{2}\sqrt{1+\tau}-\sqrt{2+\tau})(\sqrt{2+\tau}\pm
 \sqrt{2})\right]^2.
\end{equation}
As $\tau$ increases, i.e. the anisotropy of the lattice increases,
the critical value for the potential increases and the
localization becomes more difficult (Fig. \ref{fig:2}). As $t_f^\prime\rightarrow 0$, the system becomes two dimensional and there
is no threshold for localization, as expected. For large $t_f^\prime/t_f$, $V_c\sim\sqrt{t_f^\prime t_f}$, which gives
$V_c\rightarrow 0$ in the one dimensional limit, $t_f\rightarrow 0$.
Moreover, it is possible to evaluate the integral for nonzero $\epsilon$ \cite{delves}, yielding
\begin{equation}
 \tilde{V}=\frac{2(\sqrt{1-(2-\tau)^2z}+\sqrt{1-(2+\tau)^2z})}{w\left(\frac{2}{\pi}\right)^2K[k_+(\tau)]K[k_-(\tau)]},
\end{equation}
where $w=2+\tau+\epsilon/2$, $z=1/w^2$ and

\begin{eqnarray}
 k_\pm^2&=&\frac{1}{2}-\frac{1}{2}\left[\sqrt{1-(2-\tau)^2z}+\sqrt{1-(2+\tau)^2z}\right]^{-3}\\\nonumber
&\times&\left[\sqrt{1+(2-\tau)\sqrt{z}}\sqrt{1-(2+\tau)\sqrt{z}}+\sqrt{1-(2-\tau)\sqrt{z}}\sqrt{1+(2+\tau)\sqrt{z}}\right]\\\nonumber
&\times&\left\lbrace\pm 16z+\sqrt{1-\tau^2z}\left[\sqrt{1+(2-\tau)\sqrt{z}}\sqrt{1+(2+\tau)\sqrt{z}}+\sqrt{1-(2-\tau)\sqrt{z}}\sqrt{1-(2+\tau)\sqrt{z}}\right]^2\right\rbrace.
\end{eqnarray}

\begin{figure}
 \centering
 \includegraphics[scale=0.8]{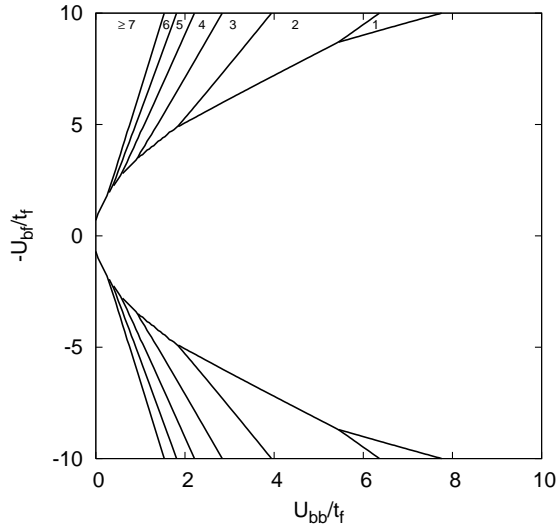}
  \caption{Phase diagram in the presence of lattice anisotropy (for $\tau =t_f^\prime/t_f= 1.5$ and $\mu = (n_0-1/2)U_{bb}$). To be compared with Fig. \ref{v_ubb_minener_01} ($\tau = 1$, $\mu = (n_0-1/2)U_{bb}$). One can see that anisotropy with $\tau>1$ causes the localization threshold to move to higher values of $|U_{bf}|$.}
 \label{v_ubb_minener_02}
\end{figure}

Using this exact result, the phase diagram can be obtained for arbitrary $\tau$. In Fig. \ref{v_ubb_minener_02}, we display the phase diagram for $\tau=1.5$. Comparing Fig. \ref{v_ubb_minener_02} with Fig. \ref{v_ubb_minener_01} (isotropic case)  we see that the phase
boundaries are closer to the $U_{bf}$ axis.

\section{Effects of higher impurity bands}\label{sec4}

An important point one always has to keep in mind about the
optical lattice experiments is that the effective Hubbard models,
such as Eq. (\ref{hamilton}), are obtained by projecting the system
into the lowest band of the lattice \cite{jaksch}. This procedure is expected to describe
the low energy physics as long as the band gaps are larger than
the temperature and interaction scales in the problem. In the
equivalent language of Wannier functions, this condition
corresponds to requiring the Wannier function of each lattice site
to be undisturbed by interactions.

\begin{figure}
 \centering
 \includegraphics[scale=0.4]{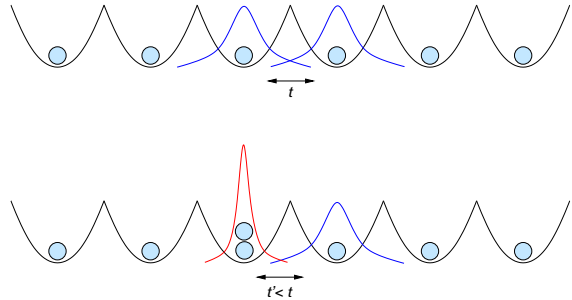}
 \caption{Schematic representation of the effect of higher impurity bands to the hopping parameter. If the localized impurity attracts extra particles (holes) to the localization site, the local wave function of the impurity particle changes. Then the hopping parameter for this site is different from that for the other sites.}
 \label{fig:7}
\end{figure}

In the context of the current problem, we discussed the
critical hopping strength that is needed to localize the impurity
particle to a small region, which is of the order of one lattice
site. The precise determination of the Hubbard model parameters
such as $U_{bf}$ depends on the microscopic model one starts from. For the Hubbard model to work correctly, the Wannier functions for the
impurity must be unchanged even if the impurity particle is
localized to one lattice site. As a localized impurity attracts
(or repels) extra particles (holes) to its localization site, one may
expect the on-site wave function of the localized particle to be
different from the Wannier functions at other lattice sites. This
is essentially considering the coupling of the localized particle
to higher impurity bands, and should be a small effect
controlled by the parameter $\frac{U_{bf}}{\Delta_f}$, where
$\Delta_f$ is the width of the first band gap of the impurity
bands. Thus, the effect we are considering in this
section would be important only if the impurity particle is highly
mobile in the lattice, while the interaction between the background
particles and the impurity is strong enough to localize the
particle (Fig. \ref{fig:7}).

In such a case, the system can still be modelled by a Hubbard model
where the hopping strength between the localization site and its
neighbors ($t'_f$) is different from the hopping strength between
any other neighboring sites in the lattice ($t_f$). These hopping strengths can once
again be calculated by looking at the overlaps of the localized
wavefunctions between neighboring lattice sites \cite{jaksch}.

\begin{figure}
 \centering
 \includegraphics[scale=0.8]{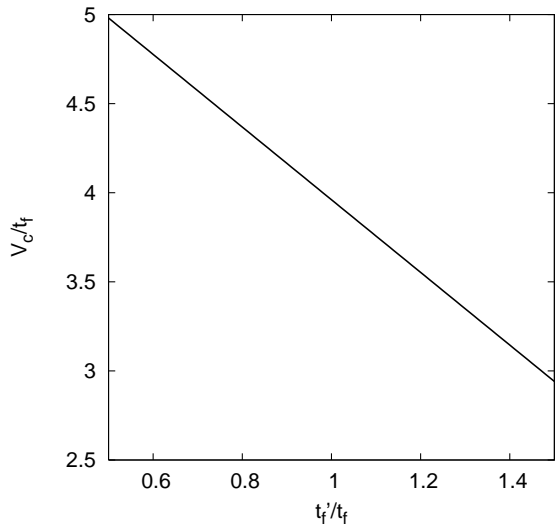}
  \caption{The critical value of interaction $V_c/t_f$ as a function of $\tau=t_f^\prime/t_f$ (Eq. (\ref{Vc3})). As the ratio of the hopping parameters $\tau$ increases, localization occurs for smaller values of the interaction.}
 \label{fig:3}
\end{figure}

\begin{figure}
 \centering
 \includegraphics[scale=0.8]{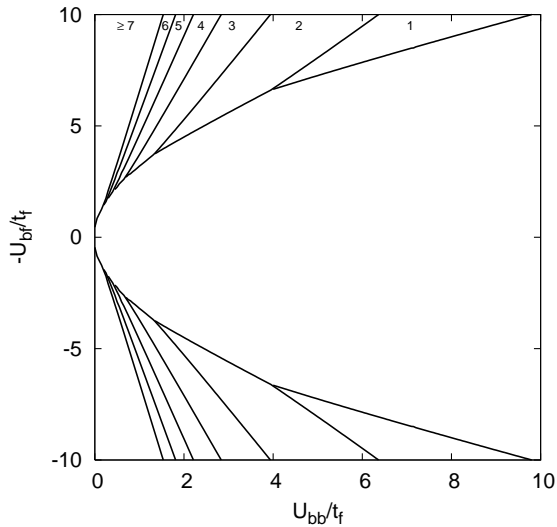}
  \caption{Phase diagram obtained when the effect of higher impurity bands is taken into account ($\tau = 1.5$, $\mu = (n_0-1/2)U_{bb}$).
  This effect is modelled by the parameter $\tau$, which is the
  ratio of the hopping strength between the localization site and its neighbors to
  the one between any other neighboring sites. Compare this figure with Fig. \ref{v_ubb_minener_01} ($\tau = 1$, $\mu = (n_0-1/2)U_{bb}$). One can see that localization is easier if $\tau>1$.}
 \label{v_ubb_minener_03}
\end{figure}

In this case, we take the single particle Hamiltonian as
\begin{equation}
 H=-t_f\sum_{< i,j>}f_i^\dagger f_j-(t_f^\prime-t_f)\sum_{< l,m>}f_l^\dagger f_m(\delta_{l0}+\delta_{m0})-Vf_0^\dagger f_0.
\end{equation}
Calculations similar to those performed in the previous sections
yield
\begin{equation}
 \tilde{V}=\frac{1-\int_{-\pi}^\pi\int_{-\pi}^\pi\int_{-\pi}^\pi\frac{d\theta_xd\theta_yd\theta_z}{(2\pi)^3}\frac{2(\tau-1)\sum_i\cos\theta_i}{6-2\sum_i\cos \theta_i+\epsilon}}{\int_{-\pi}^\pi\int_{-\pi}^\pi\int_{-\pi}^\pi\frac{d\theta_xd\theta_yd\theta_z}{(2\pi)^3}\frac{1}{6-2\sum_i\cos
 \theta_i+\epsilon}},
\end{equation}
where $\tilde{E}=-6-\epsilon$. Evaluating this integral exactly \cite{joyce2}, we obtain the relation
\begin{equation}
 \tilde{V}=\frac{2\tau}{\frac{(1-\eta)^{1/2}}{\omega}\left(1-\frac{1}{4}\eta\right)^{1/2}\left(\frac{2}{\pi}\right)^2
 K(k_+)K(k_-)}-2(\tau-1)\omega,
\end{equation}
where $\eta=-16z(\sqrt{1-z}+\sqrt{1-9z})^{-2}$,
$z=1/\omega^2=1/(3+\epsilon/2)^2$, and $k_\pm^2=\frac{1}{2}\left[1\pm
\eta\sqrt{1-\frac{1}{4}\eta}-\left(1-\frac{1}{2}\eta\right)\sqrt{1-\eta}\right]$. When $\epsilon=0$, the critical value for the potential is found to be
\begin{equation}\label{Vc3}
 \tilde{V_c}\approx 3.95678[1-0.51622(\tau-1)].
\end{equation}
As can be seen in Fig. \ref{fig:3}, the coupling to higher impurity bands can substantially change the critical value for localization.
If $U_{bf}<0$, we expect a narrowing of the local wave function (as in Fig. \ref{fig:7}), then $t_f^\prime<t_f$ and consequently localization is harder $V_c(\tau)>V_c(\tau=1)$. Similarly if $U_{bf}>0$ we expect easier localization. In general one would then expect each different polaron state to have a different $\tau$ value. Still to gain a basic understanding of this effect we obtain the phase diagram (Fig.
\ref{v_ubb_minener_03}) using constant value of $\tau=1.5$.

\section{Effects of Fluctuations of the Mott insulator}\label{sec5}

The ideal Mott insulator state is achieved only when the
boson hopping term in the Bose-Hubbard Hamiltonian is zero. When
there is a small but nonzero hopping probability for the bosons,
the ground state of the MI contains virtual particle and hole
excitations. When $t_b/U_{bb}$ gets larger, these excitations gain
more amplitude and will finally destroy the order in the MI, and
cause a transition to the SF state.

In the previous sections, we considered the MI to be a perfect
insulator by setting boson hopping to zero. In this section, we
consider the effects of small but nonzero hopping on our
calculations. We always work in the limit that the boson hopping
$t_b$ is the smallest energy scale of the system and require that
the system be away from the MI-SF transition boundaries. Under
these conditions, we can use  perturbation theory to investigate
the effects of particle and hole fluctuations on the localization
of the impurity particle.

There are two main consequences of turning on the boson hopping.
First, the compound object formed by the impurity and the extra
bosons (polaron) will become mobile.
The width of the polaron band will be proportional to $t_f
t_b^\delta$, where $\delta$ is the number of extra bosons (or
holes) forming the polaron. As $t_b$ is the smallest energy scale
in the problem, this polaron mobility effect will be small,
especially for polarons containing more than one boson or hole. In other words as $t_f\gg t_b$, the fermion is much more mobile than the boson (or the combined polaron). The mobility of the polaron will then have a small effect on the
localization problem considered above.

The second effect is the change of effective mass (or effective
hopping $t_f$) of the impurity particle due to scattering from the
fluctuations of the MI. The first contribution to such processes
will be proportional to  $(U_{bf}/U_{bb})^2$ for the fermion to scatter from a
particle-hole pair, and $(t_b/U_{bb})^2$  for the pair to be
excited. Because of the presence of $U_{bf}^2$ which is large in
the limit that we are interested in, this second effect will be
dominant over polaron mobility. We should also mention that it is
only lowest orders of the perturbation theory that these two
effects can be considered separate processes.

To the lowest order in $t_b/U_{bb}$ the boson wavefunction can be
written as: $\mid \Psi \rangle = \mid MI\rangle +
\frac{t_b}{U_{bb}}\sum_{<i,j>} b_i^\dagger b_j \mid MI\rangle$.
The state of the system without boson-fermion interaction is then
 $\mid \Psi \rangle \mid k \rangle$, where $\mid k \rangle$ is the wavefunction of a fermion moving with lattice momentum $\vec{k}$.

By treating the boson-fermion interaction as a perturbation
\begin{eqnarray}\label{V}
\hat{V} = U_{bf}\sum_i n_i f_i^\dagger f_i,
\end{eqnarray}
we can write the second order energy shift as
\begin{widetext}
\begin{eqnarray}\label{Energyshift}
\bigtriangleup ^{(2)} = \sum_{\alpha \neq \beta,
k^{\prime}}\frac{\mid \langle k^{\prime}\mid \langle MI\mid
b_{\alpha}^\dagger b_{\beta} \hat{V}\big[\mid MI\rangle +
\frac{t_b}{U_{bb}}\sum_{<i,j>} b_i^\dagger b_j \mid MI\rangle
\big]\mid k\rangle \mid ^2}{\varepsilon_k -
(U_{bb}+\varepsilon_{k^{\prime}})},
\end{eqnarray}
\end{widetext}
%
%
%
%
where the only non-vanishing term is
\begin{widetext}
\begin{eqnarray}\label{nonvanishingterm}
\langle k^{\prime}\mid \langle MI\mid b_{\alpha}^\dagger b_{\beta}
\hat{V} \frac{t_b}{U_{bb}}\sum_{<i,j>} b_i^\dagger b_j \mid
MI\rangle \mid k\rangle = \frac{t_b}{U_{bb}}\frac{U_{bf}}{N}n_0
(n_0+1)\big[e^{i(\vec{k}^{\prime}-\vec{k})\cdot
\vec{r}_{\beta}}-e^{i(\vec{k}^{\prime}-\vec{k})\cdot
\vec{r}_{\alpha}}\big].
\end{eqnarray}
\end{widetext}

Again considering the lowest non-vanishing order contribution we
get:
\begin{widetext}
\begin{eqnarray}\label{Energyshift2}
\bigtriangleup ^{(2)} &=&
\frac{t_b^2}{U_{bb}^2}\frac{U_{bf}^2}{N^2}n_0^2
(n_0+1)^2\sum_{<\alpha,\beta>,
k^{\prime}}\frac{2-2\cos[(\vec{k}^{\prime}-\vec{k})\cdot(\vec{r}_{\beta}-\vec{r}_{\alpha})]}{\varepsilon_k
- (U_{bb}+\varepsilon_{k^{\prime}})} \\ &=&
U_{bf}^2\frac{t_b^2}{U_{bb}^2}n_0^2 (n_0+1)^2
\int_{-\pi}^{\pi}\frac{d^3k^{\prime}}{(2\pi)^3}\frac{12-4\sum_{i}\cos[(\vec{k}^{\prime}-\vec{k})_i]}{2t_f
\sum_{i}[\cos(k_i^{\prime})-\cos(k_i)]-U_{bb}} \\ &\simeq&
-U_{bf}^2\frac{t_b^2}{U_{bb}^3}n_0^2 (n_0+1)^2
\bigg[12-28\frac{t_f}{U_{bb}}\sum_{i}\cos(k_i)\bigg],\:\:\frac{t_f}{U_{bb}}\!\ll\!\!1.
 \end{eqnarray}
\end{widetext}

Hence, we see that the first effect of the fluctuations of the MI
is to renormalize the hopping strength of the impurity particle to
\be t_f \rightarrow t_f \left( 1 - 14 n_0^2 (n_0+1)^2
\frac{U_{bf}^2 t_b^2}{U_{bb}^4} \right). \ee Essentially a
decreasing $t_f$ means the effective mass of the particle is
larger, and it becomes easier to localize the particle. Thus, the
effects of boson hopping on all the situations discussed in
previous sections can be obtained by scaling the phase diagrams by
the renormalized value of $t_f$.

\section{Conclusions}\label{sec6}
We study the localization of a single impurity particle in a
lattice containing a Mott insulator of bosons. This is a simple
limit of the complex physics of the Bose-Hubbard model for
mixtures of different atomic species. The impurity particle has
two distinct types of behavior; it may either move freely
throughout the lattice or may choose to localize at a certain
lattice site by attracting  extra bosons or holes. In
the limit of a perfect Mott insulator, we calculate the boundary
between these two phases as well as the number of extra
bosons (holes) forming the bound state exactly. Our result for
the phase diagram is given in Fig. \ref{v_ubb_minener_01}.

We believe that this phase diagram can be checked experimentally. In recent experiments boson-fermion
mixtures were created in the parameter regimes that we consider in
this paper \cite{gunter,ospelkaus}. If the density of fermions is small enough, one can
disregard the many-particle nature of the fermions and consider
the localization problem for one of them. To determine
whether a fermion is localized, RF spectroscopy would be an ideal
tool. The mean-field shift of a localized fermion would be larger
as it sees more background bosons on the average. Equivalently, one
can say that the binding energy of the fermion calculated in
Section \ref{sec2} would be directly reflected in the RF spectra for
fermions. Different polaron states, containing multiple numbers of
bosons or holes can be distinguished similarly.

The calculations presented in this paper were carried out with the
assumption of a homogenous infinite system. In cold gas
experiments there is always a confining trap, which in general
complicates the correspondence between the infinite system
predictions and experimental results. However, when a lattice
boson system is driven deep into the MI regime, the density
profile of the system consists of MI plateaus separated by thin SF
regions. Within each MI region, the fermion would see a flat
interaction potential, and when the fermion is localized, the width
of the wavefunction of the fermion becomes of the order of few
lattice sites. Thus, we expect our calculations to accurately
reflect the transition boundaries, especially for transitions
between different polaron states.

After the ideal case, we consider three effects which may play a
role in the experiments. The first case we consider is the
possibility of tunnelling anisotropy. As the strength of the
lattice is determined by the laser intensity (and beam waist), experiments occasionally have such anisotropy. We
generalize the exact results to this case and find that it becomes
easier to localize the impurity when the system becomes more two
dimensional, as expected. Next, we consider the
effect of higher bands of the impurity particle. We argue
that this effect can be taken into account by modifying the
hopping strengths between the localization site and its neighbors,
and obtain the phase diagram. Finally, we consider the effect of
small but finite tunnelling strength for the bosons forming the
MI, and calculate the effective hopping strength for the impurity
particle in the presence of fluctuations.

We believe that our exact results about impurity localization on a Mott
insulator background provide a starting point for the
investigation of the complex phase diagram of mixtures in optical
lattices.

\begin{acknowledgements}
S.S. is supported by TUBITAK Grant No. 106T052. R.O.U. is supported by TUBITAK. M.\"O.O. is supported by TUBA-GEBIP and TUBITAK Kariyer Grant No. 104T165.
M.\"O.O. wishes to thank Hui Zhai and B. Tanatar for useful discussions and acknowledge the hospitality of
Institut Henri Poincar\'{e}-Centre \'{E}mile Borel, where part of the work was completed.
\end{acknowledgements}

\end{document}